\documentclass[11pt]{article}
\usepackage{graphicx}
\usepackage{amsmath}
\usepackage{amssymb}
\usepackage{amsxtra}

\title{Evolution and Possible Forms of Primordial Antimatter and Dark Matter celestial objects}
\author{Maxim Yu. Khlopov\\Institute of Physics, Southern Federal University,Rostov on Don, Russia\\ Virtual Institute of Astropartcie physics, Paris, France and\\National Research Nuclear University "MEPHI", Moscow, Russia \\ email: khlopov@apc.in2p3.fr\\
O.M. Lecian\\
Sapienza University of Rome,\\
Faculty of Medicine and Pharmacy,\\
Viale Regina Elena, 324 - 00185 Rome, Italy;\\
Sapienza University of Rome,\\
Faculty of Medicine and Dentistry,\\
Piazzale Aldo Moro, 5 - 00185 Rome, Italy;\\ Kursk State University,\\ Faculty of Physics, Mathematics
and Information Sciences,\\ Chair of Algebra, Geometry and Didactics of Mathematics
Theory,\\ ul. Radis'c'eva, 33, aud. 201, Kursk, Russia. \\ email:orchideamaria.lecian@uniroma1.it}

\begin{document}
\maketitle

\begin{abstract}
The structure and evolution of Primordial Antimatter domains and Dark
matter objects are analysed. Relativistic low- density antimatter domains are described.
The Relativistic FRW perfect-fluid solution is found for the characterization of i) ultra-
high density antimatter domains, ii) high-density antimatter domains, and iii) dense anti-
matter domains. The possible sub-domains structures is analyzed. The structures evolved
to the time of galaxy formation are outlined. Comparison is given with other primordial
celestial objects. The features of antistars are outlined. In the case of WIMP dark
matter clumps, the mechanisms of their survival to the present time are discussed. The
cosmological features of neutrino clumping due to fifth force are examined.
\end{abstract}



\section{Introduction}
    The formation of antimatter regions and antimatter domains in a matter/antimatter asymmetric Universe
has long been studied according to the properties of the pertinent celestial objects, as well as to the observational
signatures expected, i.e. the energetic gamma rays descending from the matter-antimatter interaction at the 
boundaries of the antimatter domains. Several scenarios can be envisaged, i.e. also ones in which strong 
antimatter inhomogeneities interact with the surrounding medium (see \cite{PPNP,4,CKSZ,Dolgov,KRS2} for review and references).\\
The mechanisms of survival for the antimatter domains can be analyzed.\\ 
Comparison with other celestial bodies enables one to extrapolate the properties of both
the formation and evolution of such celestial bodies, as well as the interactions under which the celestial bodies
are formed.\\ 
In the present paper, low-density antimatter domains will be revised in the non-relativistic description, under the suitable hypotheses. The Relativistic diffusion equation of low-density antimatter domains will be solved; the Relativistic radius and the Relativistic spherical shell interaction width will be calculated.\\ 
Dense antimatter domains will be introduced and classified according to the density, i.e. ultra-high density antimatter domains, very-high density ones and high-density ones. The Relativistic FRW diffusion equation of dense antimatter domains will be solved in the perfect-fluid FRW plasma solution. The Relativistic radius of the dense antimatter domains in the FRW symmetry and the Relativistic spherical shell interaction width in the FRW symmetry will be calculated; the calculated expressions will be shown to depend on the Relativistic quantities in a non-trivial manner.\\ 
Baryon subdomains inside the antimatter domains will be investigated; in particular, the analysis will be conduced in the cases pertinent to the epoch before the second phase transition and that after the second phase transition. This way, the formation of non-trivial structures will be assessed; more in detail, 'Swiss-cheese' structures and 'Chinese-boxes' structures will be reconducted to the analytical quantification.\\
Antimatter-excess regions will be explored wrt the diffusion process taking place at the boundary regions.\\
The density of antimatter domains at the time of galaxy formation will be written down.\\
Experimental-verification methods will be recapitulated for antimatter domains in a matter/antimatter asymmetric Universe within the framework of inhomogeneous baryosynthesis. The investigation methods for these purposes will be specialized to the study of the $\gamma$-ray background and of the expected anti-Helium. Further experimental purposes will be recalled.\\
Comparison with other celestial bodies will be brought. The features of antimatter celestial bodies in the Galaxy, WIMP dark-matter clumps and Fifth-Force neutrino lumps will be revised for the sake of the study of the formation mechanisms, the Universe-evolution survival models, and of the interaction ruling the structure of the celestial bodies.

   \section{Low-density antimatter domain: diffusion equation}
  The Relativistic diffusion equation of 
  
   $n_{\bar{b}}$ the antibaryon number density as a function of
  $n_b$ the baryon number density, and
   $n_\gamma$ the photon number density reads \cite{kkm2000}
  \normalsize
   \begin{equation}
      \frac{dn_{\bar{b}}}{dt}=-\frac{3d}{R}<\sigma v>n_{\bar{b}}n_b-\beta n_{\bar{b}}
  \end{equation}
  being
   $R$ the antimatter domain non-Relativistic radius, and
   $d$ the antimatter domain spherical shell boundary interaction width; furthermore,
   $<\sigma v>$ antibaryon-baryon annihilation cross-section within the interaction region is defined, and
   $\beta$ the FRW Relativistic factor is introduced.\\
  
 Being
    $\bar{r}$ the antibaryon-to-photon ratio and
 $r$ baryon-to-photon ratio

the Relativistic FRW diffusion equation of low-density domains rewrites
\begin{equation}
   \bar{r}=-\frac{3d}{R}<\sigma v >rn_\gamma\bar{r}-\beta \bar{r}
\end{equation}

  \subsection{Low-density antimatter domains: non-Relativistic approximation} The non-Relativistic diffusion equation of low-density antimatter domains can be approximated after
 neglecting $\beta \bar{r}$, and
 posing $<\sigma v> rn_\gamma\Delta t\sim 1$, 

and solved as
\begin{equation}
    \frac{\bar{r}_\tau}{\bar{r}_0}=exp[-\frac{3d}{R}\int_{t_0}^{t_\tau}<\sigma v>rn_\gamma dt]
\end{equation}

\normalsize

\subsection{Low-density antimatter domains: Relativistic solution}
The Relativistic diffusion equation of low-density antimatter domains under the assumption
$\beta \bar{r}<<1$
can be solved
as
\begin{equation}
    \frac{d}{dt_\tau}(ln[\frac{r_\tau}{r_0}-\tilde{\beta}(t_\tau-t_0)])=-\frac{1}{3}\left(\frac{4\pi}{3}\right)^{1/3}\frac{\delta(t_\tau)}{a(t_\tau)^{1/3}}
\end{equation}
 with $\tilde{\beta}=-beta$.\\
The Relativistic quantities
$d\rightarrow \delta(t)$ Relativistic spherical-shell width interaction region, and
 $a(t)=\frac{4\pi R(t)^3}{3}$  FRW volume
have been upgraded.

\section{General implemetation- Relativistic\label{sectionferi}}

After hypothesizing $-\beta n_{\bar{b}}\equiv\tilde{\beta}$ small but not negligible, and $\tilde{\beta}\simeq const$, the following solution is found

\begin{equation}
ln[\frac{\bar{r}_f}{\bar{r_0}}-\tilde{\beta}\Delta t]\simeq-\frac{\delta}{3a}\Delta t
\end{equation} in the case of
perturbed Minkowski space-time.\\
In the case of an FRW symmetry, the following solution is written
\begin{equation}
\frac{d}{dt_\tau}ln[\frac{\bar{r}_f}{\bar{r_0}}-\tilde{\beta}\Delta t]\simeq -\frac{\delta(t_\tau)}{3a(t_\tau)}
\end{equation}

 \section{Perfect-fluid Relativistic FRW equation of dense antimatter domains}
 The prefect-fluid FRW diffusion equation of the antibaryon number density writes
\begin{align}
    \frac{dn_{\bar{b}}}{dt}=-\frac{3d}{R}<\sigma v>_{ext}n_{\bar{b}}n_b-\beta n_{\bar{b}}+Q(\vec{r}, p, t)-\frac{n_{\bar{b}}}{t_d}+\nonumber\\
    +\sum_iF_i(p, \dot{p}; ...)-f(\rho_E, \vec{p}; R_d, l_d; \vec{v}_T, v_f; \tilde{i})-\mu\nabla^2 n_{\bar{b}}
\end{align}
Here, 
 $<\sigma v>_{ext}$ is cross-section of the antibaryon-baryon annihilation process at the boundary of the antimatter domain,
 $Q(\vec{r}, p, t)$ is a source term (can be neglected),
$F_i(p, \dot{p}; ...)$ are further terms depending on the momentum (can be neglected),
 $f(\rho_E, \vec{p}; R_d, l_d; \vec{v}_T, v_f; \tilde{i})$ is plasma characterization in terms of the viscosity properties and of the turbulent velocity (can be neglected),
 $\frac{n_{\bar{b}}}{t_d}\simeq<\tilde{\sigma}\tilde{v}>_{int}n_{\bar{b}}n_b$ is the decay rate inside the interior of the domain, 
 $t_d=const?$ is the time scale of annihilation,
 $<\sigma v>_{int}$ is cross-section of the antibaryon-baryon annihilation process in the interior of the antimatter domain,
 $\beta n_{\bar{b}}$ accounts for the FRW homogeneous Relativistic expansion of the universe, and
 $\mu$ is the chemical potential, i.e. $\tilde{\mu}n_{\bar{b}}=-\mu\nabla^2n_{\bar{b}}$
for the self-similarity properties of the equation.
  
   \section{Dense antimatter domains}
    By construction, both the antibaryon density and the baryon one are much higher than average baryon density in all the Universe;\\ several cases can be distinguished:

$i$) {\bf ultra-high densities}\\

the antibaryon excess and baryon ones start to exceed the contribution of thermal quark-antiquark pairs before QCD phase transition\\


$ii$) {\bf very-high densities} the antibaryon density and the baryon ones exceed the contribution of plasma and radiation after the QCD phase transition\\

$iii$) {\bf high densities} the antibaryon densities and the baryon ones exceed the DM density\\

 \subsection{ $i)$ ultra-high density antimatter domains}
 Let
  $n_{\bar{b}}$ be the number density of antibaryons. The following diffusion equation is outlined
\small 
 \begin{equation}
    \frac{dn_{\bar{b}}}{dt}=-\frac{3d}{R}<\sigma v>_{ext}n_{\bar{b}}n_b-\beta n_{\bar{b}}-\mu\nabla^2 n_{\bar{b}}\equiv -\frac{n_{\bar{b}}}{t_s}-\beta n_{\bar{b}}+\tilde{\beta}+\tilde{\mu}
\end{equation}
\normalsize
and solved as
\begin{equation}
    ln[\frac{\bar{r}_\tau}{\bar{r}_0}-(\tilde{\beta}+\tilde{\mu})(t_\tau-t_0)]=-\frac{1}{3}\left(\frac{4\pi}{3}\right)^{1/3}<\sigma v>_{ext}rn_\gamma\int_{t_i}^{t_\tau}\frac{\delta(t)}{(a(t))^{1/3}}dt
\end{equation}
with
 $a(t)=4\pi R(t)^3/3$ the Relativistic FRW volume, and
 $d\rightarrow\delta(t)$ the Relativistic interaction spherical-shell width, which simplifies as

\begin{equation}
    \frac{d}{dt_\tau}(ln[\frac{\bar{r}_\tau}{\bar{r}_0}-(\tilde{\beta}+\tilde{\mu})(t_\tau-t_0)])=-\frac{1}{3}\left(\frac{4\pi}{3}\right)^{1/3}<\sigma v>_{ext}rn_\gamma\frac{\delta(t_\tau)}{(a(t_\tau))^{1/3}}
\end{equation}

\subsubsection{Relativistic expression for the radius of the antimatter domain}
Relativistic expression for the radius of the antimatter domain reads
\small
\begin{equation}
    (a(t_{\tau}))^{1/3}=-\left(\frac{3}{4\pi}\right)^{1/3}\frac{<\sigma v>_{ext}rn_\gamma\delta(t_\tau)}{3}\frac{[\frac{\bar{r}_\tau}{\bar{r}_0}-(\tilde{\beta}+\tilde{\mu})(t_\tau-t_0)]}{\frac{d}{dt_\tau}[\frac{\bar{r}_\tau}{\bar{r}_0}-(\tilde{\beta}+\tilde{\mu})(t_\tau-t_0)]}
\end{equation}
\normalsize
\subsubsection{Relativistic expression for the interaction width} 
Relativistic expression for the interaction width is obtained as
\begin{equation}
    \delta(t_\tau)\sim-\left(\frac{3}{4\pi}\right)^{1/3}\frac{3(a(t_\tau))^1/3}{<\sigma v>_{ext}rn_\gamma}\frac{\frac{d}{dt_\tau}[\frac{\bar{r}_\tau}{\bar{r}_0}-(\tilde{\beta}+\tilde{\mu})(t_\tau-t_0)]}{[\frac{r_\tau}{r_0}-(\tilde{\beta}+\tilde{\mu})(t_\tau-t_0)]}
\end{equation}
\normalsize
The relativistic expression of the interaction width of the antimatter domain depends therefore also on the Relativistic radius in a non-trivial manner, i.e. as a prefactor.

  \subsection{$ii)$ very-high-density antimatter domains}
In the case of very-high-density antimatter domains, the diffusion equation of the baryon number density becomes 
  \small
 \begin{equation}
    \frac{dn_{\bar{b}}}{dt}=-\frac{3d}{R}<\sigma v>_{ext}n_{\bar{b}}n_b-\beta n_{\bar{b}}-\frac{n_{\bar{b}}}{t_d}-\mu\nabla^2 n_{\bar{b}}\equiv -\frac{n_{\bar{b}}}{t_s}-\beta n_{\bar{b}}-\frac{n_{\bar{b}}}{t_d}-\mu\nabla^2 n_{\bar{b}}
\end{equation}
\normalsize
solved as
\small
\begin{equation}  
    ln[\frac{\bar{r}_\tau}{\bar{r}_0}+<\tilde{\sigma}\tilde{\mu}>_{int\ \ j}rn_\gamma\bar{r}-(\tilde{\beta}+\tilde{\mu})(t_\tau-t_0)]=-\frac{1}{3}\left(\frac{4\pi}{3}\right)^{1/3}\int_{t_i}^{t_\tau}\frac{\delta(t)}{(a(t))^{1/3}}dt
\end{equation}
\normalsize
with
 $a(t)=4\pi R(t)^3/3$ Relativistic FRW volume, and
 $d\rightarrow\delta(t)$ Relativistic interaction spherical-shell width
\normalsize
\normalsize
\scriptsize
\begin{equation}  
    \frac{d}{dt_\tau}(ln[\frac{\bar{r}_\tau}{\bar{r}_0}+<\tilde{\sigma}\tilde{\mu}>_{int\ \ j}rn_\gamma\bar{r}(t_\tau-t_0)-(\tilde{\beta}+\tilde{\mu})(t_\tau-t_0)])=-\frac{1}{3}\left(\frac{4\pi}{3}\right)^{1/3}\frac{\delta(t_\tau)}{(a(t_\tau))^{1/3}}
\end{equation}
\normalsize

 \subsubsection{Relativistic expression of the radius of the antimatter domain}
The Relativistic expression of the radius of very-high-density antimatter domains is obtained as

 \begin{equation}\nonumber
(a(t_\tau))^{1/3}=-\left(\frac{3}{4\pi}\right)^{1/3}\frac{<\sigma v>_{ext}rn_\gamma\delta(t_\tau)}{3}\cdot
\end{equation}
\begin{equation}
\cdot\frac{[\frac{\bar{r}_\tau}{\bar{r}_0}+<\tilde{\sigma}\tilde{\mu}>_{int\ \ j}rn_\gamma\bar{r}(t_\tau-t_0)-(\tilde{\beta}+\tilde{\mu})(t_\tau-t_0)]}{\frac{d}{dt_\tau}[\frac{\bar{r}_\tau}{\bar{r}_0}-+<\tilde{\sigma}\tilde{\mu}>_{int\ \ j}rn_\gamma\bar{r}(t_\tau-t_0)-(\tilde{\beta}+\tilde{\mu})(t_\tau-t_0)]}
\end{equation}

\subsubsection{Relativistic expression of the spherical shell interaction width}
The Relativistic expression of the spherical shell interaction width of very-high density antimatter domains is
\tiny
\begin{equation}
    \delta(t_\tau)\sim-\left(\frac{3}{4\pi}\right)^{1/3}\frac{3(a(t_\tau))^1/3}{<\sigma v>_{ext}rn_\gamma}\frac{\frac{d}{dt_\tau}[\frac{\bar{r}_\tau}{\bar{r}_0}+<\tilde{\sigma}\tilde{\mu}>_{int\ \ j}rn_\gamma\bar{r}(t_\tau-t_0)-(\tilde{\beta}+\tilde{\mu})(t_\tau-t_0)]}{[\frac{\bar{r}_\tau}{\bar{r}_0}+<\tilde{\sigma}\tilde{\mu}>_{int\ \ j}rn_\gamma\bar{r}(t_\tau-t_0)-(\tilde{\beta}+\tilde{\mu})(t_\tau-t_0)]}
\end{equation}
\normalsize
The Relativistic expression of the interaction width of the antimatter domain depends therefore also on the Relativistic radius in a non-trivial manner, i.e. as a prefactor.\normalsize    
   
   \subsection{$iii)$ high-density antimatter domains}
   The diffusion equation of the antibaryon number density of high-density antimatter domains is characterized as 

\begin{equation}\nonumber
 \frac{dn_{\bar{b}}}{dt}=-\frac{3d}{R}<\sigma v>_{ext}n_{\bar{b}}n_b-\frac{n_{\bar{b}}}{t_d}-\mu\nabla^2 n_{\bar{b}}\equiv
 \end{equation}
\begin{equation}\equiv-\frac{3d}{R}<\sigma v>_{ext}n_{\bar{b}}n_b-\frac{n_{\bar{b}}}{t_s}-\frac{n_{\bar{b}}}{t_d}-\mu\nabla^2 n_{\bar{b}}
    \end{equation}
\normalsize\vspace{2mm}
and solved as
\begin{equation}\label{eq100}
    ln[\frac{\bar{r}_\tau}{\bar{r}_0}-(\tilde{\mu})(t_\tau-t_0)]=-\frac{1}{3}\left(\frac{4\pi}{3}\right)^{1/3}<\sigma v>_{ext}rn_\gamma\int_{t_i}^{t_\tau}\frac{\delta(t)}{(a(t))^{1/3}}dt
\end{equation}
Here,
 $a(t)=4\pi R(t)^3/3$ is the Relativistic FRW volume, and
$d\rightarrow\delta(t)$ is the Relativistic interaction spherical-shell width.\\
Eq. (\ref{eq100}) rewrites
\begin{equation}  
    \frac{d}{dt_\tau}(ln[\frac{\bar{r}_\tau}{\bar{r}_0}-(\tilde{\mu})(t_\tau-t_0)])=-\frac{1}{3}\left(\frac{4\pi}{3}\right)^{1/3}<\sigma v>_{ext}rn_\gamma\frac{\delta(t_\tau)}{(a(t_\tau))^{1/3}} 
\end{equation}

 \subsubsection{Relativistic expression for the radius of the antimatter domain}
 In the case of high-density antimatter domains, the Relativistic expression for the radius of the antimatter domain is expressed as
\begin{equation}  
    (a(t_\tau))^{1/3}=-\left(\frac{3}{4\pi}\right)^{1/3}\frac{<\sigma v>_{ext}rn_\gamma\delta(t_\tau)}{3}\frac{[\frac{\bar{r}_\tau}{\bar{r}_0}-\tilde{\mu}(t_\tau-t_0)]}{\frac{d}{dt_\tau}[\frac{\bar{r}_\tau}{\bar{r}_0}-\tilde{\mu}(t_\tau-t_0)]}
\end{equation}
\subsubsection{Relativistic expression for the spherical shell interaction width}
The Relativistic expression for the spherical shell interaction width of high-density antimatter domains is solved as
\begin{equation}  
    \delta(t_\tau)\sim-\left(\frac{3}{4\pi}\right)^{1/3}\frac{3(a(t_\tau))^1/3}{<\sigma v>_{ext}rn_\gamma}\frac{\frac{d}{dt_\tau}[\frac{\bar{r}_\tau}{\bar{r}_0}-\tilde{\mu}(t_\tau-t_0)]}{[\frac{\bar{r}_\tau}{\bar{r}_0}-\tilde{\mu}(t_\tau-t_0)]}
\end{equation}
 The Relativistic expression of the spherical-shell interaction width of the antimatter domain depends therefore also on the Relativistic radius in a non-trivial manner, i.e. as a prefactor.    
   
    \section{Conditions and evolution of different types of of strong primordial inhomogeneities in non-homogeneous baryosynthesis}
In the case of non-homogeneous primordial baryosynthesis, various types of scenarios can accomplish:
 antimatter consisting of axion-like particles; closed walls for baryogenesis with excess of antibaryons, and phase fluctuations such that a baryon excess is created everywhere and with non-homogeneous distribution.
To avoid large-scale fluctuations, the fluctuations have to be imposed to be small.\\ In the latter cases of small fluctuations, the following inequality holds
\begin{equation}  
    \frac{3B}{4\pi R(t)^3}>>\rho_B
\end{equation}
The diffusion process of the model is described as follows.\\

   Three Regions can be outlined:
   
  $1$) the {\bf dense antimatter domain} of radius $R\le R_1$
  of antibaryon number density $n_{{b}\ \ 1}$, and of chemical potential $\mu_1$,
    
   $2$) the {\bf outer spherical shell} region of radius $R_1\le R\le R_2$ of antibaryon number density $n_{{b}\ \ 2}$, and  of chemical potential $\mu_2$,\\ $\ \ $where the diffusion process happens, and $3$) the outmost region of radius $R_3\ge R_2$ \small of antibaryon number density $n_{{b}\ \ 3}$  of low antimatter density.\\
 \\
 The chemical potentials of related to the three regions are assumed to be small but not negligible, i.e.
 
 $\mid \tilde{\mu_1}\mid <<1$, $\mid \tilde{\mu_2}\mid<<1$, and $\mid \tilde{\mu_3}\mid<<1$.
The differential equation of the antibaryon number density in
 Region $1$) is
\begin{equation}\label{eq200}
    \frac{n_{b\ \ 1}}{dt}=-\mu_1\nabla^2n_{b\ \ 1}\sim \tilde{\mu_1}n_{b\ \ 1};
\end{equation}\label{eq201}
The differential equation of the antibaryon number density in Region $2$) is
\begin{equation}\label{eq202}
    \frac{n_{b\ \ 2}}{dt}=-\mu_2\nabla^2n_{b\ \ 2}\sim \tilde{\mu_2}n_{b\ \ 2}
\end{equation}
The differential equation of the antibaryon number density in Region $3$)
\begin{equation}  
    \frac{n_{b\ \ 3}}{dt}=-\mu_3\nabla^2n_{b\ \ 3}\sim \tilde{\mu_3}n_{b\ \ 3}
\end{equation}
The solutions of Eq. (\ref{eq200}), Eq. (\ref{eq201}) and (\ref{eq202})

 must satisfy the  continuity conditions
\begin{equation}
    n_{b\ \ 1}(t, R_1)=n_{b\ \ 2}(t, R_1)
\end{equation}
\normalsize
on the boundary of Region $1$), and
\begin{equation}
    n_{b\ \ 2}(t, R_2)=n_{b\ \ 3}(t, R_2)
\end{equation}
on the boundary of Region $2$).

 \section{Dense Baryon subdomains}
It is possible to hypothesize 
the presence of antibaryons inside the baryon subdomains, which exceed the survival size of volume $V_j=4\pi R_j^3/3$; such a possibilty is dependent on the second phase transition.\\
For axion-like particles, it is dependent on the QCD phase transition.\\
Two possibilities are outlined, i.e. according to whether the description is taken before the $\Lambda_{QCD}$ phase transition, or after it.\\
I) In the case $\Lambda<\Lambda_{QCD}$, 
  the baryon number density $n_b$ in a baryon subdomain filled with (grazing) antibaryons obey the following plasma characterization
\begin{equation}  
    \frac{dn_b}{dt}=-<\sigma v>_{j\ \ ext}n_bn_{\bar{b}}-<\sigma v>_{j\ \ int}n_bn_{\bar{b}}-\mu\nabla^2n_b
\end{equation}
The following perfect-fluid Relativistic FRW solution is found

\begin{equation}\nonumber
    -\sqrt{4\pi}{3}\frac{\tilde{\delta}(t_\tau)}{3\tilde{a}(t_\tau)}r_{int}\bar{r}_{int}n_\gamma<\sigma v>_{j\ \ ext}=
    \end{equation}\begin{equation}=\frac{d}{dt_\tau}ln[r_{int}\bar{r}_{int}n_{\gamma\ \ int}<\sigma v>_{j\ \ ext}+\\-\tilde{\mu}(t_\tau-t_0)+]
\end{equation}

In the case II) $\Lambda>\Lambda_{QCD}$
the baryon number density $n_b$ in a baryon subdomain 
without free antibaryons inside\\
is described by the following plasma characterization
\begin{equation}
    \frac{dn_b}{dt}=-<\sigma v>_{j\ \ ext}n_bn_{\bar{b}}-\mu\nabla^2n_b
\end{equation}
The following perfect-fluid Relativistic FRW solution is found
\begin{equation}
    -\sqrt{4\pi}{3}\frac{\tilde{\delta(t_\tau)}}{3\tilde{a}(t_\tau)}r_{int}\bar{r}_{int}n_\gamma<\sigma v>_{j\ \ ext}=\frac{d}{dt_\tau}ln[+\tilde{\mu}(t_\tau-t_0)]
    \normalsize
\end{equation}  
   \section{Further structures}
   Further structures can be analysed, according to the presence of baryon subdomain(s) inside the antibaryon domain.\\
\subsection{'Swiss-cheese' structures} 
A description of 'Swiss-cheese' structures can be hypothesized as an antimatter domain containing one matter domain, in the simplest instance, and more complicated 'Swiss-cheese' structures, such as an antimatter domain containing several matter subdomains.

\subsubsection{Antibaryon domain containing one baryon subdomain}
In the case of an antibaryon domain containing one baryon subdomain, the antibaryon number density obeys the differential equation

\begin{equation}\nonumber
\frac{dn_{\bar{b}}}{dt}=-\frac{3d}{R}<\tilde{\sigma} \tilde{v}>_{ext}n_{\bar{b}}n_b-\beta n_{\bar{b}}+Q(\vec{r}, p, t)-\frac{n_{\bar{b}}}{t_d}+\\+F_i(p, \dot{p}; ...)+\end{equation}\begin{equation}-\mu\nabla^2 n_{\bar{b}}-\frac{3d_i}{R_i}<\hat{\sigma}_i\hat{v}_i>n_{\bar{b}}n_{bi}-\mu_i\nabla^2n_{\bar{b}}
\end{equation}

\normalsize
\subsubsection{Swiss-cheese structure: baryon domain containing several baryon subdomains}
\begin{equation}\nonumber\frac{dn_{\bar{b}}}{dt}=-\frac{3d}{R}<\tilde{\sigma} \tilde{v}>_{ext}n_{\bar{b}}n_b-\beta n_{\bar{b}}+Q(\vec{r}, p, t)-\frac{n_{\bar{b}}}{t_d}+\end{equation}\begin{equation}+\sum_{i=1}^{i=I}\left(F_i(p, \dot{p}; ...)-\mu\nabla^2 n_{\bar{b}}-\frac{3d_i}{R_i}<\hat{\sigma}_i\hat{v}_i>n_{\bar{b}}n_{bi}-\mu_i\nabla^2n_{\bar{b}}\right)
\end{equation}
\normalsize\\
\subsubsection{Chinese-boxes structures}
'Chinese-boxes' structures are described as
\begin{equation}\nonumber
  \frac{dn_{\bar{b}}}{dt}=-\frac{3d}{R}<\sigma v>_{ext}n_{\bar{b}}n_b-\beta n_{\bar{b}}+Q(\vec{r}, p, t)-\frac{n_{\bar{b}}}{t_d}-\mu\nabla^2n_{\bar{b}}\end{equation}\begin{equation}\nonumber-\sum_{i=1}^{i=I}\left[F_i(p, \dot{p}; ...)-\mu\nabla^2 n_{\bar{b}_i}-\frac{3d_i}{R_i}<\sigma_v>_{i\ \ ext}n_{\bar{b}}n_{b_i}\right]+\sum_{j=1}^{j=J}[F_j(p, \dot{p}; ...)\end{equation}\begin{equation}-\mu\nabla^2 n_{\bar{b}_j}-\frac{3d_j}{R_j}<\sigma_v>_{j\ \ ext}n_{\bar{b}}n_{b_j}]
  \end{equation}
  
     \section{Galaxy formation: Relativistic density of the surviving domains}
     The present section is aimed at studying the density of the antimatter domains at the time of galaxy formation.
     
     \subsection{Plasma characterization}
     The plasma characterization of the antimatter domains at the time of galaxy formation is given after the condition
\begin{equation}  
    <\sigma v>_{int}r=0,
\end{equation}
i.e. after the antibaryon/baryon interactions in the interior of the antimatter domains have exhausted.\\
The following conditions are taken into account:
\begin{equation}
    \tilde{\mu}(t_\tau-t_0)n_\gamma<<1,
\end{equation}
and
\begin{equation}
    \tilde{\beta}(t_\tau-t_0)n_\gamma<<1,
\end{equation}
with
\begin{equation}
    (\tilde{\mu}+\tilde{\beta})(t_\tau-t_0)n_\gamma<<1,
\end{equation}
\normalsize
i.e. that the chemical-potential etrms and the Relativistic FWR terms be small but not negligible.  
  
\subsubsection{$i)$ ultra-high-density antimatter domains}
In the case of ultra-high-density antimatter domains, the antimatter-domain density at the time of galaxy formation reads
\begin{equation}\nonumber
\frac{\bar{r}_\tau n_{\gamma}}{a(t_\tau)}=\frac{1}{a(t_\tau)}\frac{n_{\gamma}}{\frac{1}{\bar{r}_0}+(\tilde{\beta}+\tilde{\mu})(t_\tau-t_0)n_{\gamma}}\cdot\end{equation}\begin{equation}\cdot exp\left[{\frac{1}{3}}\left(\frac{4\pi}{3}\right)^{1/3}<\sigma v>_{ext}r_0n_\gamma\int_{t_0}^{t_\tau}\frac{\delta{t}}{a(t)}dt\right]
\end{equation}
\subsubsection{$ii)$ very-high-density antimatter domains}
In the case of very-high density antimatter domains, the antimatter-domain density at the time of galaxy formation is
\begin{equation}\nonumber
\frac{\bar{r}_\tau n_{\gamma}}{a(\tau)}=\frac{1}{a(t_\tau)}\frac{n_\gamma}{\frac{1}{\bar{r}_0n_\gamma}+(\tilde{\beta}+\tilde{\mu})(t_\tau-t_0)n_{\gamma}}\cdot\end{equation}\begin{equation}exp^{\frac{1}{3}}(\frac{4\pi}{3}^{1/3})<\sigma v>_{ext}r_0n_\gamma\int_{t_0}^{t_\tau}\frac{\delta{t}}{a(t)}dt
\end{equation}
\subsubsection{$iii)$ high-density antimatter domains}
In the case of high-density antimatter domains, the antimatter-domain density at the time of galaxy formation becomes
\small
\begin{equation}  
\frac{\bar{r}_\tau n_{\gamma}}{a(\tau)}=\frac{1}{a(t_\tau)}\frac{n_\gamma}{\frac{n_{\gamma}}{\bar{r}_0}+\tilde{\mu}(t_\tau-t_0)n_{\gamma}}e^{\frac{1}{3}}(\frac{4\pi}{3}^{1/3})<\sigma v>_{ext}r_0n_\gamma\int_{t_0}^{t_\tau}\frac{\delta{t}}{a(t)}dt
\end{equation} 
\normalsize
  
   \section{Experimental verification}
   The signatures of the experimental verification of the existence of antimatter domains have to be analysed. In particular, the $\gamma$-ray background is expected to be modified after the baryon/antibaryon interaction within the boundary interaction region of the antimatter domains. Furthermore, the detection of anti-Helium flux after the $AMS2$ experiment is awaited.\\
The properties of $pp$ atoms have been studied in \cite{{aea1985}}
\normalsize
In \cite{aea1986}, the $\gamma$-ray spectrum originated after the $p\bar{p}$ annihilation in liquid Hydrogen is analysed by means of two spectrometers. As a result,
no exotic narrow peacks are evidentiated, and the upper limit is calculated.\\
\normalsize

The $\gamma$-ray signal due to matter-antimatter annihilation on the boundary of an antimatter domain can therefore be analyzed \cite{cdg1998}; the hypotheses of a matter/antimatter symmetric Universe and of a matter/antimatter asymmetric Universe can be scrutinized and compared.\\ 
In the case of a matter/antimatter-symmetric Universe:
more $\gamma$-rays than the observed quantities are predicted; therefore,
 a matter/antimatter-symmetric Universe is possible iff the present Universe is one consisting of the matter quantity.\\
\normalsize
\normalsize

The $p\bar{p}$ interaction process is studied as resolving in photons after the $\pi^0$ decay. Be

 $\bar{g}$ the mean photon multiplicity;

 each $p\bar{p}$ annihilation process is estimated to produce $\bar{g}\simeq3.8$ electrons and positrons, and\
an approximate similar number of photons.\\
The annihilation electrons are described at a redshift $y$ s.t. $20<y<1100$.\\
The mechanisms that control the electrons motion must therefore be studied. Such mechanisms are evaluated to be
the cosmological redshift, the collision with CBR photons, and
the collision with ambient plasma electrons.\\
At the considered values of the redshift,
the collisions with CBR photons are considered the most important control mechanism of the electron trajectory.\\
For initially-Relativistic electrons of energy $E_0=\gamma_0 m_e$,  
the dependece of the width of the reheated zone where the electrons produced after the annihilations directly deposit energy into the fluid, i.e. the
electron range, on $\gamma_0$ is negligible.\\
The inclusive photon spectrum in the $p\bar{p}$ process is normalized to $\bar{g}$; the average number of photons
made per unit volume is calculated:
the transport equation of the photons scatter and redshift, (which lead to a spectral flux of annihilation photons), is therefore assessed. A conservative lower limit for the $\gamma$-ray signal can this way estimated.\\    
  
 The $\gamma$-rays energy
is expected to be of
 order $100 MeV-10 Mev$ at modern times; a different value can be expected for the opaque universe at the early stages).\\
The results are awaited after the experiment
 $AMS2$ as far as the presence of the anti-Helium flux is concerned.

\section{Further experimental verifications}
Further experimental verifications of a matter/antimatter Universe can be expected .\\
\\
As an example \cite{psc2004}, annihilation and
 transformation of annihilating matter’s rest mass into energy particles and radiation with $100\% $ efficiency can be looked for at different length scales.\\
A substantial lack of antimatter on the Earth is evidentiated within the due limits.\\
A lack of antimatter in the vicinity of the Earth is found.\\
Matter asymmetry in the Solar System can be revealed within the study of the:\\  Solar wind, i.e. the continuous outflow of
particles from the Sun. For antiplanets of radius $r$ and distance $d$ from the Earth intercepting the Solar wind, the expected annihilation flux is\\
$F(\gamma=100MeV)\sim10^8(r/d)^2\ \ photons\ \ cm^{-2}\ \ s^{-1}$.\\
At scales larger than the Solar System, i.e. at the Galaxy scales,
the $\gamma$-ray analysis must be investigated ($\gamma$-ray detectors have better detection capabilities and localization ones at $E\sim 100\ \ MeV$ than neutrino
detectors).
Antimatter mixed in with matter inside our Galaxy’s gas at $E\sim 100 \ \ MeV$ is expected to be present in a matter/antimatter-symmetric Universe.\\
\normalsize
\normalsize
Models can be postulated \cite{dki2003}, such that SUSY-condensate baryogenesis models motivate the possibilities of antimatter domains in the Universe.\\
In this case, vast antimatter structures in Early-Universe evolution possible after
 initial space distribution at the inflationary
stage of the quantum fluctuation field $\phi(r, t_0)$, unharmonic potential
of the field carrying the baryon charge, and
inflationary expansion of the initially microscopic
baryon distribution. The vast antimatter regions are calculated to be
separated at distances larger than $10\ \ Mpc$ from the Earth, and separated from the matter ones by baryonically-empty voids.\\
Such models are not ruled out after\\
 cosmic rays data,
$\gamma$-rays ones, and CMB anisotropy ones.\\
  
Antimatter in a matter/antimatter-symmetric Universe can be further verified \cite{adr1997} after the presence of antimatter at the Galactic scale and above.\\
As far as hydrogen in ''clouds'' is concerned, the experimental verification is based on the observation of $\gamma$-rays from their directions, compatible with $\pi^0$ decay, and
 non-observation of a $\gamma$ excess. In this case, form the observational data, the  antibaryon presence in the media is calculated not to exceed one part in $10^{15}$.\\
The instance of galaxy-antigalaxy collisions can be studied. Such events have not been verified after devices s.t. Antennae pair NGC4038(9). Clusters of certain galaxies, dense enough and active in order to allow for intergalactic hot plasma in the central
parts (at temperatures of order $\sim 10\ \  keV$:
it is therefore possible to verify the  presence of antimatter as a few parts per million from the observation of absence of enough $\gamma$-ray excess on the thermal spectral tail.\\

Large antimatter regions with sizes larger than the critical surviving size can be verified in different observational proofs \cite{cds2012}. The absence of anti-Helium in the cosmic rays and annihilation signals can be consistent as an indicator: their fraction in the Galaxy is smaller than $10^{-4}$. The antimatter islands must be separated from a space filled
with matter at least by the distance of about $1 Mpc$.\\
For this, the possibility
for antimatter islands (antistars) in the Galaxy still allowed \cite{dnv2013}.\\
\normalsize
Large antimatter regions with high antimatter density evolve to single galaxies \cite{gru2015}. They are detected after particular content of  anti-Helium and anti-deuterium.\\
\normalsize
  
Further Cosmic antimatter searches can be pursued \cite{cas2007}. The presence of antistars in
our Galaxy can be verified after the
 possibility to detect antinuclei with $Z\ge2$.  Domain sizes of the scale of galaxies or scales of galaxy clusters can be testified
after antimatter cosmic
rays (CR) originating from the nearest domain for
 uniform domains,
non-uniform domains, and
condensed antimatter bodies (i.e. antistars, antiplanetoids).
The upper limit of antistars in the Milky Way has been estimated as $10^7$
(i.e. $10^{-4}$ of the total
 number of stars).
Antistars can be described as confined
into compact structures separated from the matter environment and able to survive for a longer period rather than in gas clouds.
Antistars are not expected to be strong $\gamma$-ray emitters, unless they at least cross
a galactic cloud or impact on other condensed  bodies.\\
\normalsize

The lower limit on the distance of the nearest antistar \cite{dwo1994} has been set as $\sim 30\ \  pc$. The
 upper limit on the fraction of antistars in the Andromeda Galaxy has been estimated as $\sim10^{-3}$.\\
\normalsize
  
Experimental verification of presence of matter regions and antimatter ones in a matter/antimatter-symmetric universe should be studied after the pre-recombinational signals and the post-recombinational ones.\\ 
The prerecombination signal \cite{ad21998} allows one for the verification of
the presence of domains of larger size. The assumption that matter domains and antimatter ones were in contact before the last scattering exhibits such effects after which
contact and annihilation significantly distort the
radiation from the last-scattering surface: a single domain
boundary, or a fraction , can be detectable; differently, the
 absence of such signatures rules out a matter/antimatter-symmetric universe.\\

The postrecombination signal \cite{ad21998} would consists of the observable unobserved $\gamma$-ray flux, due to nuclear annihilation rate of matter/antimatter near domain boundaries; the a resulting relic diffuse $\gamma$-ray flux exceeds the observed cosmic diffuse $\gamma$ spectrum, so that a matter/antimatter-symmetric Universe is ruled out unless the matter region consists of almost the entire Universe\\
\normalsize
   \section{Antistars}
The analysis of the mean free path of the cross section of the matter/antimatter annihilation products in the interaction spherical shell boundary of the antistars is conisistent for the comparison with the $\gamma$-ray-background constraints \cite{dol2021}.
  
After the compilation of the 10-years Fermi Large Area Telescope (LAT) $\gamma$-ray-sources catalog, 
 constraints on the abundance of antistars around the Sun are obtained:
 14 antistar candidates are present around the Sun. In particular, they have been chosen as they are
 not associated with any objects belonging to established $\gamma$-ray source classes, and
exhibit a spectrum compatible with baryon-antibaryon annihilation \cite{dtv2021}.

\normalsize

\section{Antimatter celestial objects in the Galaxy}
The exist observational evidences of the existence of antimatter celestial objects in the Milky Way; more in detail \cite{bdo2007}, they are
 point-like sources of gamma radiation, and
 diffuse galactic $\gamma$-ray background, where the latter 
 possible antimatter sources are to be verified after an anomalous abundance
of chemical (anti-)elements around it possibly measured by spectroscopy,
 anti-nuclei in cosmic rays, and
 more exotic events, where large amounts of matter
and antimatter interact. In the latter case,
 star-antistar
annihilation can be considered: huge energy produced, even though their total destruction is prevented by the radiation pressure produced in the collision; and
collision of a star and an anti-star with
similar masses is calculated to provoke a peculiar result.\\ 
  
\subsection{Antistars}
The creation of stellar-like objects in the very early universe \cite{dbl2014}, from the QCD phase transition until the BBN and later, can be witnessed as the presence of
 some of the celestial objects created which can consist of antimatter.
The $\alpha$ cosmological baryon asymmetry $\alpha=\frac{N_B}{N_\gamma}$  can be close to unity, i.e. much larger than the observed
value $\alpha\simeq6\cdot10^{-10}$. The ratio $\alpha$ can also be
 negative: this way, the
amount
of antimatter constituting compact objects in the
Galaxy is expected.\\
  
 \section{WIMP's clumps}
\subsection{\bf Neutralino clumps}
Within the 
standard cosmological scenario (FRW with its thermal history, inflationary-produced primordial fluctuation spectrum and with a hierarchical clustering), the neutralino clumps \cite{bde2008} undergo
 tidal destruction in the hierarchical clustering (i.e. the smaller
clumps are captured by the larger clumps) at early stages of the structure-formation process, starting from a time of clump detachment from the Universe expansion.\\
In the case of small-scale dark-matter clumps, a mass function can be calculated for the survived clumps:
the  tidal destruction of clumps by the Galactic disk,
the life-time
of clumps in the central stellar bulge, and
the life-time of clumps in the stellar halo
spheroid can be calculated; as a result,
 the minimal mass is the evaluated as the  Moon-scale mass.\\
\normalsize
  
\subsection{Neutralino annihilation in the Galaxy}
Within the
standard cosmological scenario,
 neutralino annihilation of small-scale neutralino clumps \cite{bde2006}
would produce a signal from the  galactic halo: the
 clump destruction is due to larger-scale clumps,
 gravitational field of the galactic disk,
stars in the galactic bulge, and
stars in the galactic halo.\\
The mutual tidal clumps interactions would become important at
 early stages of hierarchical clustering, and for the galactic halo formation.\\ The
 hierarchical clustering implies
 clumps surviving the hierarchical clustering to be continuously destroyed by  interactions with the galactic disk and stars. This way,

 $20\%$ of neutralino clumps surviving the hierarchical clustering between the Earth and the Moon can 'survive the Sun position' because of tidal destruction due to Galactic disk. Furthermore,
the diminishing of the expected DM annihilation signal from the galactic halo would be awaited.\\
\subsection{Small-scale DM clumps}
The clumps scenarios comprehend spherical models,
non-spherical models, and
clumps around topological defects \cite{bde2014a}.\\
The possible observational verifications are established DM-particles direct detection, record of clumps in gravitational-wave detectors,
 neutralino stars,
 baryons in clumps, and
 clump motion in the Sky sphere.\\
  
   \section{Fifth-Force neutrino lumps}
\subsection{Fifth-Force codifications}
The Fifth Force potential can be codified as \cite{ngp2005}, \cite{hju2009}\\
   \small
  \begin{equation}V(r)=\frac{\tilde{G}m_1}{r}\left(1+\alpha e^{-r/\lambda}\right)=G\frac{m_1m_2}{r}\left(1+e^{-\beta\frac{\phi}{r}}\right)= \frac{m_1 m_2}{r}\left(G+\frac{G}{\Delta G}\right).\end{equation}
  \normalsize
Such a codification allows for the description of dark-matter gravitational clustering.\\
  
\subsection{Modellizations for neutrino cosmology}
The parameter $\beta$ is intended as the Fifth-Force parameter, and
 $\phi-\nu$ coupling is postulated.\\
 The fifth force is requested to be subdominant with respect to the gravitational force \cite{cpw2016}, \cite{aww2013}. As an example, the request can be expressed as
 \begin{equation}\beta=-\frac{d \ln m_\nu}{d\phi}.\end{equation}
Is is also possibe to set a $\lambda$ comoving length scale
larger than the typical lump sizes, but
smaller than their
typical distances, such that the 
mean distance between neighboring
lumps be of order $100h^{-1}Mpc$.\\
$l$ lumps of masses $M_l$
are expressed via smoothed fields $\hat{\phi}$. The
effective coupling
\begin{equation}\beta_l=-\frac{d\ln M_l}{d\hat{\phi}}\end{equation}
is worked out.\\

\subsection{Applications for fluids of composite objects}
Neutrino lumps are described within a hydrodynamic framework, i.e. endowed with a balance equation \cite{aww2013}, and
a stability equation \cite{bbe2009} based on the Tolman-Oppenheimer-Volkoff equation.\\

    
  Within the framework of the  $\phi-\nu$ coupled fluid,
    neutrino fluctuations are hypothesized to grow under the effect of the  Fifth-Force \cite{bpa2011}.\\
   Non-Relativistic neutrino clusters under the effect of the fifth-force
are hypothesized at scales estimated to be around a few 10–$100 Mpc$.
A statistical distribution of neutrino lumps is expressed as a function of
the mass
 at different redshifts $z\ge1$.\\
The oscillating structure formation is described as at the time a large number of neutrinos  were staying in
 gravitationally-bounded lumps at $z = 1.3$.\\

\subsection{\bf Formation of large-scale neutrino lumps in a recent cosmological epoch}
Within the framework of a $\phi-\nu$ interaction,
the non-linear features of the Fifth-Force can be outlined \cite{nsw2011}.\\
The averaged interaction strength $<\beta>$ of the neutrinos in a neutrino lump reads
\begin{equation}<\beta>=-M\frac{d\ln<m_\nu>}{d\phi}.\end{equation}
The effective suppression of the $\phi$- mediated attractive force between 
neutrino lumps is proportional to $2\beta$. In particular,
the attraction between two equal
lumps is reduced by a factor $\left(\frac{<\beta>}{\beta}\right)^2$. Furthermore, the characteristic time scale for the infall
increased by a factor $\frac{\beta}{<\beta>}$ compared to the consideration excluding non-linear effects and thus results in a
slow down of the infall:
the  time scale
for the clumping of lumps to larger lumps enhanced by
a factor $\left(\frac{\beta}{<\beta>}\right)^2$.\\
\normalsize
In the interior of the lump, the possibility of a  time variation of fundamental constants results much smaller than the cosmological evolution; therefore, it is
possible to reconcile the cosmological variations of the fine structure constant with  geophysical bounds.
  
\subsection{CMB verification for  neutrino lumps}
Within the framework of a $\phi-\nu$ interaction,
the integrated Sachs-Wolfe effect of CMB \cite{pwa2010} can be considered.
The size of the gravitational potential induced by the neutrino lumps, and
the time evolution of the gravitational potential induced by the neutrino lumps have to be analyzed.\\
as a result, 
a  proportionality between the
scalar potential and the neutrino-induced gravitational
potential is found as 
\begin{equation}\beta\delta\phi=2\beta^2\Phi_\nu\end{equation}
 for the local potential and the cosmological-averaged potential.\\
The population of lumps of size $\ge$  $100 Mpc$ can lead to observable effects from the CMB anisotropies for low angular momenta.\\

 \section{Outlook and perspectives}
Evolution of antimatter domains have been studied: an analysis of
 low-density antimatter domains and
 dense antimatter domains has been performed. More in detail,
 ultra-high density antimatter domains,
very-high density antimatter domains, and
 high-density antimatter domains.\\
\vspace{2mm}
Experimental verification of their signatures consists of the search for confirmation in
the observed $\gamma$-ray background, and
for the expected anti-Helium flux in $AMS02$ experiment.\\
  Comparison with other celestial objects has been accomplished: study of
 formation mechanisms,
 Universe-evolution survival models, and
 comparison of interactions characterizing the structure of the celestial bodies has been performed.

\section*{Acknowledgements}
OML acknowledges the Programme Education in Russian Federation for Foreign Nationals of the Ministry of Science and Higher
Education of the Russian Federation. The work by MK was performed in MEPHI in the framework of cosmological studies of Prioritet2030 programme.


\end{document}